\documentclass[pra,twocolumn,showpacs,floatfix]{revtex4}

\usepackage{amssymb}
\usepackage{amsmath}
\usepackage{graphicx}

\def\be{\begin{equation}}
\def\ee{\end{equation}}
\def\e#1{\label{#1}\end{equation}}
\def\bea{\begin{eqnarray}}
\def\eea{\end{eqnarray}}
\def\ea#1{\label{#1}\end{eqnarray}}

\def\bem#1{\begin{mathletters}\label{#1}}
\def\eml{\end{mathletters}}

\def\4#1{{\boldsymbol{#1}}}
\def\8#1{{\widetilde{#1}}}
\def\bse{\begin{subequations}}
\def\ese{\end{subequations}}

\begin{document}

\title{Optimal Dynamical Decoherence Control of a Qubit}
\author{Goren Gordon}
\author{Gershon Kurizki}
\affiliation{Department of Chemical Physics, Weizmann Institute of Science, Rehovot
76100, Israel }
\author{Daniel A. Lidar}
\affiliation{Departments of Chemistry, Electrical Engineering, and Physics, Center for
Quantum Information Science \& Technology, University of Southern
California, Los Angeles, California 90089, USA}

\begin{abstract}
A theory of dynamical control by modulation for optimal decoherence reduction is developed. 
It is based on the non-Markovian Euler-Lagrange equation for the energy-constrained
field that minimizes the average dephasing rate of a qubit for any
given dephasing spectrum.
\end{abstract}


\maketitle

\textit{Introduction}.--- Quantum information processing (QIP) harbors
enormous unleashed potential in the form of efficient algorithms for
classically intractable tasks and unconditionally secure cryptography
\cite{Jaeger:book}. Perhaps the largest hurdle on the way to a
realization of this potential is the problem of decoherence, which
results when a quantum system, such as a quantum computer, interacts
with an uncontrollable environment or bath
\cite{Breuer:book}. Decoherence reduces the information processing
capabilities of quantum computers to the point where they can be
efficiently simulated on a classical computer.  In spite of dramatic
progress in the form of a theory of fault tolerant quantum error
correction (QEC) \cite{Gaitan:book}, finding methods for overcoming
decoherence that are both efficient and practical remains an important
challenge. An alternative to QEC that is substantially less
resource-intensive is dynamical decoupling (DD)
\cite{Viola:98Zanardi:98bViola:99Vitali:99,KhodjastehLidar:04,Uhrig:07}.
In DD one applies a succession of short and strong pulses to the
system, designed to stroboscopically decouple it from the
environment. This can significantly slow down decoherence, though not
halt it completely, since unlike QEC, DD does not contain an entropy
removal mechanism.  Similar in spirit to DD (in the sense of being
feedback-free), but more general, is the method we term here
\textquotedblleft dynamical control by modulation\textquotedblright\
(DCM), wherein one may apply to the system a sequence of
arbitrarily-shaped pulses whose duration may vary anywhere from the
stroboscopic limit to that of continuous dynamical modulation
\cite{Kofman:04,Agarwal:99,Alicki:01a,Facchi:05,Brion:05,gor06a}. In
the DCM approach, the decoherence rate is governed by a universal
expression, in the form of an overlap between the bath-response and
modulation spectra, subject to finite spectral bandwidth and amplitude
constraints.

Neither DD
\cite{Viola:98Zanardi:98bViola:99Vitali:99,KhodjastehLidar:04,Uhrig:07}
nor DCM \cite{Kofman:04,gor06a} studies have so far gone beyond
particular schemes for suppression of decoherence. What is lacking is
a systematic theory for finding the \emph{optimal modulation} for any
given decoherence process. Here we apply variational principles to the
DCM approach in order to address this problem.  We derive an equation
for the \emph{optimal, energy-constrained control by modulation}
(ODCM) that minimizes dephasing, for any given dephasing spectrum. We
numerically solve this equation, and compare the optimal modulation to
energy-constrained DD pulses. We show that ODCM always outperforms DD
when subjected to the same energy constraint. We note that
Ref.~\cite{Uhrig:07} developed an optimal DD pulse sequence for the
diagonal spin-boson model of pure dephasing, but without an energy
constraint, i.e., assuming zero-width pulses. This was improved upon
by perturbatively accounting for pulse widths in
Ref.~\cite{Karbach:08}.

\textit{Model}.--- We consider a driven two-level system (qubit) with
ground and excited states ${|g\rangle }$ and ${|e\rangle }$ separated
by energy $\omega _{a}$ (we set $\hbar =1$), and 
Hamiltonian
\begin{equation}
H(t)=\left( \omega _{a}+\delta _{r}(t)\right) {|e\rangle }{\langle e|}
+\left( V(t){|g\rangle \langle e|}+h.c.\right) ,  \label{Hamiltonian}
\end{equation}
where $V(t)=\Omega(t)e^{-i\omega _{a}t}+c.c.$ is a time-dependent
resonant classical driving 
field with amplitude $\Omega (t)$, and $\delta _{r}(\omega )$ describes
random, Gaussian distributed, zero-mean energy fluctuations. Let ${|\psi
(t)\rangle }$ denote the solution of the time-dependent Schr\"{o}dinger
equation with the Hamiltonian $H(t)$, and let the density matrix $\rho (t)=
\overline{{|\psi (t)\rangle \langle \psi (t)|}}$ denote the corresponding
ensemble average over realizations of $\delta _{r}(t)$. We are interested in
the average fidelity ${\langle F(t)\rangle }$, where ${\langle \cdots
\rangle }$ is the average over all possible initial pure states of the
fidelity $F(t)=|{\langle \psi (0)|\rho (t)|\psi (0)\rangle }|$. It can be
shown that \cite{gor06a}: 
\begin{eqnarray}
{\langle F(t)\rangle } &=&1-\alpha R(t)t  \label{R-def} \\
R(t) &=&2\mathrm{Re}\left\langle \int_{0}^{t_{1}}dt_{2}\Phi
(t_{1}-t_{2})\epsilon ^{\ast }(t_{1})\epsilon (t_{2})\right\rangle
_{t}^{t_{1}} \\
\Phi (t) &=&\overline{\delta _{r}(t)\delta _{r}(0)}\quad \epsilon
(t)=e^{-i\int_{0}^{t}dt_{1}\Omega (t_{1})}  \label{eps-def}
\end{eqnarray}
where $0<\alpha \lesssim 1$ is a known constant, $\langle \cdot \rangle
_{t}^{t_{1}}\equiv \frac{1}{t}\int_{0}^{t}\cdot dt_{1}$ is the time-average, 
$R(t)$ is the \emph{average modified dephasing rate}, $\Phi (t)$ is the
second ensemble-average moment of the random (stationary non-Markov)
noise, and $\epsilon (t)$ is the phase factor associated with the modulation.

We impose the energy bound constraint
\begin{equation}
\int_{0}^{T}dt\left\vert \Omega (t)\right\vert ^{2}=E  \label{energy-const}
\end{equation}
where $T$ is the total modulation time and $E$ is the energy constraint. As
a boundary condition we require that the field is turned on, i.e. $\Omega
(0)=0$.

Although the analysis below is given in the time-domain, it is advantageous
to analyze the problem in the frequency domain, in terms of the
universal expressions \cite{Kofman:04,gor06a}: 
\begin{eqnarray}
R(t) &=&2\pi \int_{-\infty }^{\infty }d\omega G(\omega )F_{t}(\omega )
\label{R-omega} \\
G(\omega ) &=&(2\pi )^{-1}\int_{-\infty }^{\infty }dt\Phi (t)e^{i\omega t} \\
F_{t}(\omega ) &=&|\epsilon _{t}(\omega )|^{2}/t\quad \epsilon _{t}(\omega )=
\frac{1}{\sqrt{2\pi }}\int_{0}^{t}dt_{1}\epsilon (t_{1})e^{i\omega t_{1}}
\label{Ft}
\end{eqnarray}
where $G(\omega )$ is the dephasing spectrum, $\epsilon _{t}(\omega )$ is
the finite-time Fourier transform of the modulation function, and $
F_{t}(\omega )$ is the normalized spectral modulation intensity.

The model we have just described applies to a qubit
undergoing dephasing due to coupling to a finite-temperature bath of
harmonic oscillators with energies $\hbar \omega _{\lambda }$. The qubit then
has an average modified dephasing rate of the form given by
Eqs.~\eqref{R-def},\eqref{R-omega} where the dephasing spectrum is
given by \cite{gor06a}:
\begin{eqnarray}
G(\omega ) &=&\left( n(\omega )+1\right) G_{0}(\omega )+n(-\omega
)G_{0}(-\omega ) \\
G_{0}(\omega ) &=&\sum_{\lambda }|\kappa _{\lambda }|^{2}\delta (\omega
-\omega _{\lambda })
\end{eqnarray}
where $G_{0}(\omega )$ is the zero-temperature bath spectrum, $\kappa
_{\lambda }$ is the off-diagonal coupling coefficient of the qubit to the bath
oscillator $\lambda $, and $n(\omega )=\left( e^{\beta \omega }-1\right)
^{-1}$ is the average number of quanta in the oscillator (bath mode) with
frequency $\omega $, with $\beta $ the inverse temperature.

Since $R(t)$ is the overlap between the dephasing and modulation
spectra, it can be reduced by choosing an appropriate modulation that
reduces this overlap \cite{Kofman:04,gor06a,Alicki:01a}.  We shall
show that the optimal modulation reduces the spectral overlap of the
dephasing and modulation spectra (Fig.~\ref{Fig-2}).  However, since
the energy constraint in the frequency domain is non-trivial we shall
derive the equations for optimal modulation using the time domain.

\textit{Optimization}.--- We wish to find the optimal modulation, i.e.,
time-dependent near-resonant field, that minimizes $R(t)$. Calculus of variation
is an often-used technique of optimal control theory, e.g.,
\cite{PhysRevA.37.4950,Palao:02}.
We apply it to derive the Euler-Lagrange (EL)
equations for the energy-constrained optimal modulation. The accumulated
phase due to the modulation is $\phi (t)=\int_{0}^{t}d\tau \Omega (\tau )$.
Let us write $\Phi (t)=\tilde{\Phi}(t)e^{i\Delta t}$, where $\tilde{\Phi}
(t)$ and $\Delta $ are the amplitude and spectral center of the correlation
function, respectively. Using Eqs.~\eqref{R-def} and \eqref{energy-const},
we can then derive the EL equation for the optimal modulation (see
Appendix~\ref{appA} and \ref{appB}): 
\begin{eqnarray}
\lambda \ddot{\phi}(t)&=&-Z[t,\phi (t)]  \label{Z-def} \\
Z[t,\phi (t)] &=&\left\langle \tilde{\Phi}(\left\vert t-t_{1}\right\vert
)\sin [\phi (t)-\phi (t_{1})+\Delta (t-t_{1})]\right\rangle
_{T}^{t_{1}}, \notag
\end{eqnarray}
where $\lambda $ is the Lagrange multiplier. The boundary conditions for the
accumulated phase are $\phi (0)=\dot\phi (0)=0$, which results in a
smooth solution and accounts for turning the control field
on. Eliminating $\lambda $ we find that the optimal control field
shape is the solution to the following equation (see Appendix~\ref{appB}):
\begin{equation}
\ddot{\phi}(t)=\frac{-\sqrt{E}Z[t,\phi (t)]}{\sqrt{\int_{0}^{T}dt_{1}\left
\vert \int_{0}^{t_{1}}dt_{2}Z[t_{2},\phi (t_{2})]\right\vert ^{2}}}.
\label{phi-eq}
\end{equation}

Equation (\ref{phi-eq}) is the central result of this work. It
furnishes the optimal time-dependent field amplitude, that maximizes the
average fidelity ${\langle F(t)\rangle }$ at the final time $T$, via $\Omega
(t)=\dot{\phi}(t)$. Although Eq.~(\ref{phi-eq}) is a complicated
non-linear integro-differential equation, it is very useful indeed, as
we show next.

\textit{Linearized EL equation}.--- Assuming that we have a good initial
guess $\phi _{0}(t)$ for the modulation, we can look for the optimal
\emph{deviation} $\nu (t)$ by writing $\phi (t)=\phi _{0}(t)+\nu (t)$,
where $\nu (t)\ll 1$. Expanding Eq.~(\ref{Z-def})\ in powers of $\nu (t)$ and retaining
only the first order, the linearized EL equation becomes (see Appendix~\ref{appC}): 
\begin{eqnarray}
&&\lambda \ddot{\nu}(t)+\left\langle Q(t,t_{1};\phi _{0}(t))\left( \nu (t)-\nu
(t_{1})\right) \right\rangle _{T}^{t_{1}} =-C(t;\phi _{0},\lambda )
 \notag \\
 &&Q(t,t_{1};\phi _{0}(t)) = \tilde{\Phi}(\left\vert t-t_{1}\right\vert ) \times \notag \\
 &&\qquad \cos
\left( \phi _{0}(t)-\phi _{0}(t_{1})+\Delta (t-t_{1})\right) \notag \\
&&C(t;\phi _{0}(t),\lambda ) = \lambda \ddot\phi _{0}(t)+Z[t,\phi
  _{0}(t)].
\label{Q-line-def}
\end{eqnarray}
This linearized EL equation is valid also in the case of short time optimal
modulation, for which we simply set $\phi _{0}(t)=0$, subject to $\nu (t)\ll
1$ for $0\leq t\leq T\ll 1$.

\textit{Numerical analysis}.--- Armed with the equations for the optimal
modulation, we turn to solving them numerically for specific decoherence
scenarios, defined by their dephasing spectra $G(\omega )$. We obtain the
numerical solution to the integro-differential Eq.~\eqref{phi-eq} via an
iterative process, where we guess a probable solution that satisfies the
boundary conditions and the constraint, use it in the RHS of Eq.~
\eqref{phi-eq} to compute the integral, and solve the resulting
differential equation. The solution is then used in the RHS of Eq.~
\eqref{phi-eq}, and so on. 

For the examples presented below, we checked that several
initial guesses converged to the same optimal modulation. Most importantly, we 
found that the optimal modulation is {\em robust against random
control field imperfections}. This is due to the fact that the decoherence rate is determined by the
accumulated phase and not the instantaneous modulation, Eq.~\eqref{eps-def}.
Specifically, we found that a $10\%$ zero-mean random pulse fluctuation results in
less than a $1\%$ increase in the optimal dephasing rate.

We compare the optimal dephasing rate to the one obtained by the popular
periodic DD control (``bang bang'') procedure
\cite{Viola:98Zanardi:98bViola:99Vitali:99}, but to make the
comparison meaningful we impose the same energy 
constraint. Finite-duration periodic DD against pure dephasing is the ``bang bang''
application of $n$ $\pi $-pulses and is given in our setting by 
\begin{equation}
\Omega (t)=\left\{ 
\begin{array}{ll}
\pi /\nu & j\tau \leq t<j\tau +\nu \quad j=0\ldots n-1 \\ 
0 & \mathrm{otherwise}
\end{array}
\right.  \label{BB-def}
\end{equation}
where $\nu <\tau $ is the width of each pulse and $\tau $ is the interval
between pulses. The energy constraint $E$ and the total modulation duration $
T=n\tau +\nu $ are related via $n=\nu E/\pi ^{2}$. In the frequency domain,
the spectral modulation intensity can be described by a series of peaks,
where the two main peaks are at $\pm \pi /\tau $. Thus, the peaks are
shifted in proportion to the energy invested in the modulation. However, DD
is not an admissible solution to our EL equation due to its discontinuous
derivative. In order to improve the comparison, we apply our linearized
EL equation with the DD modulation as an initial guess, and obtain the
optimal modulation in the vicinity of the DD control.

\begin{figure}[th]
\centering
\includegraphics[width=9cm]{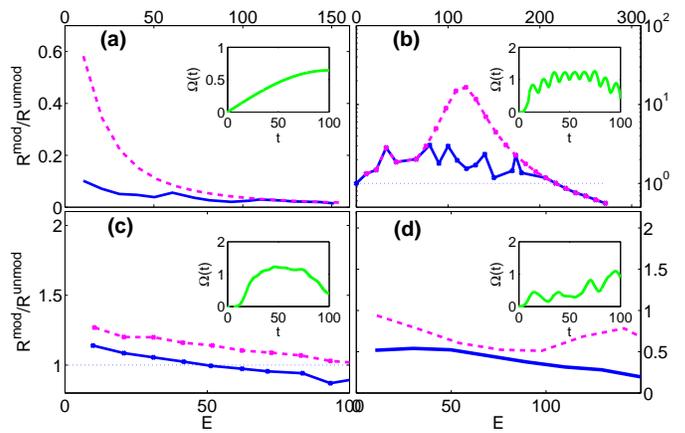}
\caption{(color online) Average modified final decoherence rate $R(T)$, normalized with
respect to the unmodulated rate, as a function of energy
constraint. DD - dash, magenta. Optimal modulation - solid, blue. 
Insets: optimal modulation $\Omega (t)$ for different energy constraints.
(a) Single-peak resonant dephasing spectrum (inset: $E=20$). (b) Single-peak
off-resonant spectrum (inset: $E=50$). (c) $1/f$ spectrum (inset: $E=30$).
(d) Multi-peaked spectrum (inset: $E=30$). }
\label{Fig-1}
\end{figure}

\begin{figure}[th]
\centering
\includegraphics[width=9cm]{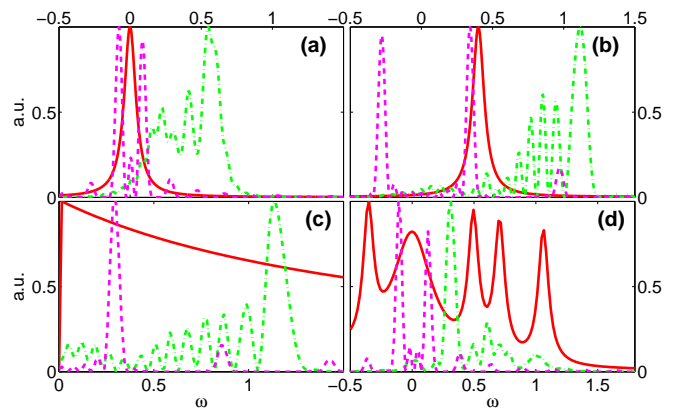}
\caption{(color online) Dephasing spectrum $G(\protect\omega )$
  (solid, red), optimal (dot-dash, green) and
DD (dash, magenta) modulation spectra $F_{T}(\protect\omega )$, in
  arbitrary units (a.u.).
Same parameters as in the insets of Fig.~\protect\ref{Fig-1}. }
\label{Fig-2}
\end{figure}

\textit{(a) Single-peak resonant dephasing spectrum}.-- This simple dephasing
spectrum describes a common scenario where $\Phi (t)=e^{-t/t_{c}}\gamma
/t_{c}$, where $\gamma $ is the long-time dephasing rate [$R(t\rightarrow\infty )=2\pi
\gamma $] and $t_{c}$ is the noise correlation time. Fig.~\ref{Fig-1}(a)
shows $R(T)$, normalized to the bare (unmodulated) dephasing rate, as a
function of the energy constraint. As expected, the more energy is available
for modulation, the lower is the dephasing rate. For low energies the
optimal modulation significantly outperforms DD, while at higher energies
this difference disappears. These results can be understood from
Fig.~\ref{Fig-2}(a), by noticing that the two central DD peaks have
significant 
overlap with $G(\omega )$ at the low energy value shown. As $E$ is increased
at fixed $T$ the DD peaks move farther apart, and have less overlap
with $G(\omega )$, leading to improved 
performance. Applying the linearized EL equation with DD as initial guess
yields only mild improvements (not shown).
The explanation for the superior performance of the optimal modulation is
also evident from Fig.~\ref{Fig-2}(a):\ since higher frequencies have lower
coupling strength in this case, the optimal control \textquotedblleft
reshapes\textquotedblright\ so as to maximize its weight in the
high-frequency range, to the extent permitted by the energy constraint.
The modulation can be well approximated by
$\Omega(t)=a[1+e^{-t/T}(t/T-1)]$, where $a$ is determined by the
energy constraint, which fits the inset in Fig.~\ref{Fig-1}(a).

\textit{(b) Single-peak off-resonant dephasing spectrum}.--- This dephasing
spectrum describes a variation on the aforementioned scenario, where the
spectral peak is shifted [$\Delta \neq 0$ in $\Phi (t)=\tilde{\Phi}
(t)e^{i\Delta t}$], e.g., coupling to a non-resonant bath. With no other
constraints, the optimal modulation is trivially similar to the one of the
resonant spectrum, with a shifted energy-constraint
$E^{\mathrm{non-res}}=E^{\mathrm{res}}+\Delta $. However, by imposing
a positivity constraint, $\dot\phi (t)\geq 0$ (positive field amplitude),
one obtains non-trivial behavior of $R(T)$ as a
function of the energy constraint -- see Fig.~\ref{Fig-1}(b). Here we used
the linearized EL equation with the DD modulation (\ref{BB-def}) as an
initial guess. For both the DD and optimal modulations, we observe an
initial \emph{increase} in the dephasing rate as a function of energy,
followed by a decrease. For DD, this can be interpreted as a manifestation
of the initial anti-Zeno effect and the subsequent quantum Zeno effect
\cite{Kofman:04,Facchi:05}. Because of the positivity constraint,
the optimal modulation does worse than the unmodulated case, for low enough
energy. The DD modulation is optimal for small energy constraints,
hence the decoherence rates of DD and our 
optimal solution coincide. This is because the DD peaks do not overlap the
off-resonant spectral peak. However, as the positive-frequency main DD peak
[Fig.~\ref{Fig-2}(b)] nears the off-resonant spectral peak, with increased
energy, the optimal modulation diverges from the DD modulation, and
\textquotedblleft reshapes\textquotedblright\ itself so as to couple to
higher modes of the bath. In the time domain [Fig.~\ref{Fig-1}(b) inset],
this is seen as a smoothing of the
abrupt DD modulation. At even higher energy constraints, there is once more
no improvement by the optimal modulation over DD, yet there is an improvement over the
unmodulated case. Over the entire range of $E$,
the optimal modulation results in a much flatter $R(T)$ than DD, which is an
indicator of its robustness. While DD is strongly influenced by the
off-resonant peak, the optimal modulation exploits the energy available to
find the minimal overlap, irrespective of dephasing spectrum.

\textit{(c) $1/f$ dephasing spectrum}.--- The ubiquitous $1/f$ dephasing
spectrum that describes a variety of experiments --
e.g., charge noise in superconducting qubits \cite{Nakamura:02} -- is
given in our notation by $G(\omega )\propto 1/\omega $, with cutoffs $\omega
_{\mathrm{min}}$ and $\omega _{\mathrm{max}}$. Fig.~\ref{Fig-1}(c)
shows that as expected, the more energy 
is available for modulation, the lower the dephasing rate. Since, as
in case (a), higher frequencies now have lower coupling
strength, the optimal control \textquotedblleft reshapes\textquotedblright\
so as to have as high a weight in the high frequency range as the
energy constraint allows [Fig.~\ref{Fig-2}(c)]. This is expressed in
the time-domain [Fig.~\ref{Fig-1}(c) inset] as the initial increase in
the modulation strength ($t<50$). The 
later decrease in modulation strength can be attributed to the lower cutoff,
where the optimal modulation benefits
from lower frequencies, i.e., lower modulation amplitudes. Upon comparing the 
$1/f$ case to the Lorentzian spectrum, Fig.~\ref{Fig-1}(a), we observe a
similar optimal initial chirped modulation in the time domain.
Despite the differences in the long-time behavior (due to the lower cutoff
in the $1/f$ case), these two examples allow us to generalize to any
dephasing spectrum with a monotonically decreasing system-bath coupling strength as a
function of frequency. The optimal modulation for such spectra will be an
energy-constrained chirped modulation, with variations due to other
spectral characteristics, e.g., cutoffs.

\textit{(d) Multi-peaked dephasing spectrum}.--- This describes the most general
scenario, where there can be several resonances and noise correlation times.
Fig.~\ref{Fig-1}(d) shows $R(T)$ as a function of the energy constraint.
Once again, because DD does not account for the dephasing spectrum, its
performance is much worse than the optimal modulation, whose
\textquotedblleft reshaping\textquotedblright\ results in monotonically
improving performance: the peaks of the optimal modulation are
predominantly anti-correlated with the peaks of $G(\omega)$.

\textit{Conclusions}.--- We have found the optimal modulation for
countering pure dephasing upon imposing an energy constraint on the
DCM approach \cite{Kofman:04,gor06a}, by deriving and solving the
Euler-Lagrange equation~\eqref{phi-eq}.  This yields optimal reduction
of the overlap of the dephasing and the modulation intensity
spectra. We stress that our optimal control theory results are also
applicable to scenarios other than pure dephasing, such as amplitude
noise (relaxation), due to the universality of
Eqs.~(\ref{R-def})-(\ref{Ft}) \cite{Kofman:04,gor06a}. The form of the
energy constraint will then differ in detail from the pure dephasing case.
However, our general conclusions about the optimal modulation to
minimize spectral overlap, will remain valid.  We expect that the
optimal modulation technique will find useful applications in quantum
information processing and quantum computation. The price is that one
must acquire intimate knowledge of the noise spectrum, which is often
neglected, as previous control techniques such as DD and QEC had no
use for it. We have shown that this information can result in the
maximization of fidelity, under operational constraints.

{\it Acknowledgments}.--- G.K. acknowledges the support of GIF and EC
(MIDAS STREP, FET Open). D.A.L. was sponsored by the United States
Department of Defense and supported under grant NSF CCF-0523675.


\newpage

\appendix

\begin{widetext}
\section{General comments on deriving optimal functions}
\label{appA}

For the optimal control of a functional 
\begin{equation}
\mathcal{F}(y,\dot y)=\int_{0}^{T}dtF(t,y,\dot y)
\end{equation}%
with the constraint 
\begin{equation}
\mathcal{K}(y,\dot y)=\int_{0}^{T}dtK(t,y,\dot y)=E
\end{equation}%
one follows the following procedure:

(i) Solve the Euler-Lagrange equation: 
\begin{equation}
\frac{\delta F}{\delta y}-\frac{\partial }{\partial t}\frac{\delta F}{\delta
\do y}=-\lambda \left[ \frac{\delta K}{\delta y}-\frac{\partial }{%
\partial t}\frac{\delta K}{\delta \dot y}\right]   \label{eq-1}
\end{equation}%
where $\delta F$ is the variation of $F$, $\lambda $ is the Lagrange
multiplier, and the boundary conditions are $y(0)=y_{0}$ and $\dot y(0)=y_{1}$.

(ii) Insert the solution $\tilde{y}(t;\lambda )$ into the constraint: 
\begin{equation}
\mathcal{K}(\tilde{y}(t;\lambda ),\tilde{\dot y}(t;\lambda ))=E
\label{eq-2}
\end{equation}%
and obtain $\lambda =\lambda (E)$.

(iii) Eliminate $\lambda $ by inserting $\lambda (E)$ into $\tilde{y}%
(t;\lambda )$ and obtain the optimal solution, $\tilde{y}(t;E)$, that
minimizes the functional $\mathcal{F}$, under the constraint $\mathcal{K}=E$.

\section{Derivation of the Euler-Lagrange equation}
\label{appB}

The average modified decoherence rate is given by: 
\begin{eqnarray}
R(T) &=&\frac{2}{T}\int_{0}^{T}dt\int_{0}^{t}dt_{1}\tilde{\Phi}(t-t_{1})\cos
[\phi (t)-\phi (t_{1})+\Delta (t-t_{1})] \\
&=&\frac{2}{T}\int_{0}^{T}dt\int_{0}^{T}dt_{1}\Theta (t-t_{1})\tilde{\Phi}%
(t-t_{1})\cos [\phi (t)-\phi (t_{1})+\Delta (t-t_{1})]
\end{eqnarray}%
where $\Theta (t)$ is the Heaviside step function.

One arrives at the following variation of the average modified decoherence
rate: 
\begin{eqnarray}
\delta R(T) &=&\frac{2}{T}\int_{0}^{T}dt\int_{0}^{T}dt_{1}  \nonumber
\label{d-J-3} \\
&&\left[ -\Theta (t-t_{1})\tilde{\Phi}(t-t_{1})\sin [\phi (t)-\phi
(t_{1})+\Delta (t-t_{1})]\delta \phi (t)\right.   \nonumber \\
&&\left. +\Theta (t-t_{1})\tilde{\Phi}(t-t_{1})\sin [\phi (t)-\phi
(t_{1})+\Delta (t-t_{1})]\delta \phi (t_{1})\right]   \label{d-R-1} \\
&=&\frac{2}{T}\int_{0}^{T}dt\int_{0}^{T}dt_{1}  \nonumber \\
&&\left[ -\Theta (t-t_{1})\tilde{\Phi}(t-t_{1})\sin [\phi (t)-\phi
(t_{1})+\Delta (t-t_{1})]\delta \phi (t)\right.   \nonumber \\
&&\left. -\Theta (t_{1}-t)\tilde{\Phi}(t_{1}-t)\sin [\phi (t)-\phi
(t_{1})+\Delta (t-t_{1})]\delta \phi (t)\right]   \label{d-R-2} \\
&=&\frac{2}{T}\int_{0}^{T}dt\int_{0}^{T}dt_{1}\left[ -\tilde{\Phi}%
(\left\vert t-t_{1}\right\vert )\sin [\phi (t)-\phi (t_{1})+\Delta
(t-t_{1})]\delta \phi (t)\right]   \label{R-var}
\end{eqnarray}%
where we have made a $t\leftrightarrow t_{1}$ substitution in the
second integrand of
Eq.~(\ref{d-R-1}), and notice that $\Theta
(t)f(t)+\Theta (-t)f(-t)=f(\left\vert t\right\vert )$.

One can easily see that defining the constraint functional as: 
\begin{equation}
  \mathcal{K}(t,\phi (t),\dot \phi(t))=\int_{0}^{t}dt_{1}|\dot
  \phi(t_{1})|^{2}=E,
  \label{const-functional}
\end{equation}%
with $K(t,\phi (t),\dot \phi(t))=|\dot \phi (t_{1})|^{2}$ results
in the variation: 
\begin{equation}
\delta K=2\ddot\phi(t)\delta \dot\phi(t).
\label{const-var}
\end{equation}%
Combining Eqs.~(\ref{R-var}), (\ref{const-var}) and (\ref{eq-1}) results in
the Euler-Lagrange equation:
\begin{equation}
\lambda \ddot\phi(t)+Z[t,\phi (t)]=0  \label{EL}
\end{equation}%
where 
\begin{equation}
Z[t,\phi (t)]=\frac{1}{T}\int_{0}^{T}dt_{1}\tilde{\Phi}(\left\vert
t-t_{1}\right\vert )\sin [\phi (t)-\phi (t_{1})+\Delta (t-t_{1})].
\end{equation}

\section{Derivation of the linearized Euler-Lagrange equations}
\label{appC}

In some cases it is advantageous to linearize the EL equations with respect
to the modulation. If one looks for the optimal deviation $\nu (t)$ from a
given pulse shape, $\phi _{0}(t)$, then one can write $\phi (t)=\phi
_{0}(t)+\nu (t)$, where $\nu (t)\ll 1$. 
Equation~(\ref{EL}) then becomes:
\bea
\lambda(\ddot\phi_0(t)+\ddot\nu(t))&=-&
\frac{1}{T}\int_{0}^{T}dt_{1}\tilde{\Phi}(\left\vert
t-t_{1}\right\vert )
\Big[\sin [\phi_0 (t)-\phi_0 (t_{1})+\Delta (t-t_{1})]\cos[\nu(t)-\nu(t_{1})]\nonumber\\&&
+\cos [\phi_0 (t)-\phi_0 (t_{1})+\Delta (t-t_{1})]\sin[\nu(t)-\nu(t_{1})]\Big]\\
&=-&\frac{1}{T}\int_{0}^{T}dt_{1}\tilde{\Phi}(\left\vert
t-t_{1}\right\vert )
\Big[\sin [\phi_0 (t)-\phi_0 (t_{1})+\Delta (t-t_{1})]\nonumber\\&&
+\cos [\phi_0 (t)-\phi_0 (t_{1})+\Delta (t-t_{1})][\nu(t)-\nu(t_{1})]\Big]+O(\nu^2(t))
\eea
where we approximated $\sin[\nu(t)-\nu(t_{1})]\approx
\nu(t)-\nu(t_{1})+O(\nu^3(t))$ and $\cos[\nu(t)-\nu(t_{1})]\approx
1-\frac{1}{2}(\nu(t)-\nu(t_{1}))^2 +O(\nu^4(t))$. 
The linearized Euler-Lagrange
equation becomes: 
\begin{equation}
\lambda \ddot\nu(t)+\frac{1}{T}\int_{0}^{T}dt_{1}Q(t,t_{1};\phi
_{0}(t))\left( \nu (t)-\nu (t_{1})\right) =-C(t;\phi _{0},\lambda )
\label{EL-line-def}
\end{equation}%
where 
\begin{eqnarray}
Q(t,t_{1};\phi _{0}(t)) &=&\tilde{\Phi}(\left\vert t-t_{1}\right\vert )\cos
\left( \phi _{0}(t)-\phi _{0}(t_{1})+\Delta (t-t_{1})\right) 
\\
C(t;\phi _{0}(t),\lambda ) &=&\lambda \ddot\phi _{0}(t)+Z[t,\phi
_{0}(t)].
\end{eqnarray}

\end{widetext}

\end{document}